\documentclass[aps,pre,twocolumn,groupedaddress,showpacs,showkeys,amsmath,amssymb]{revtex4-1}
\usepackage{graphicx}% Include figure files
\usepackage{dcolumn}% Align table columns on decimal point
\usepackage{bm}

\begin{document}

\preprint{APS/123-QED}

% Use the \preprint command to place your local institutional report
% number in the upper righthand corner of the title page in preprint mode.
% Multiple \preprint commands are allowed.
% Use the 'preprintnumbers' class option to override journal defaults
% to display numbers if necessary
%\preprint{}

%Title of paper
\title{Influence of primary particle density in the morphology of agglomerates}

% repeat the \author .. \affiliation  etc. as needed
% \email, \thanks, \homepage, \altaffiliation all apply to the current
% author. Explanatory text should go in the []'s, actual e-mail
% address or url should go in the {}'s for \email and \homepage.
% Please use the appropriate macro foreach each type of information

% \affiliation command applies to all authors since the last
% \affiliation command. The \affiliation command should follow the
% other information
% \affiliation can be followed by \email, \homepage, \thanks as well.
\author{M. D. Camejo}
\email{mdcamejo@ing.uc3m.es}
%\homepage[]{Your web page}
%\thanks{mdcamejo@ing.uc3m.es}
\author{D. R. Espeso}
\email{david.rodriguez.espeso@uc3m.es}
\author{L. L. Bonilla}
\altaffiliation[Corresponding author,]{ bonilla@ing.uc3m.es}
%\thanks{}
%\altaffiliation{}
\affiliation{G. Mill\'an Institute of Fluid Dynamics, Nanoscience and\\ Industrial Mathematics,\\ {\scriptsize \textit{Universidad Carlos III de Madrid,}}\\
{\scriptsize \textit{Avenida de la Universidad 30, 28911 Legan\'es, Spain. Fax: +34 91 624 91 29}}}

%Collaboration name if desired (requires use of superscriptaddress
%option in \documentclass). \noaffiliation is required (may also be
%used with the \author command).
%\collaboration can be followed by \email, \homepage, \thanks as well.
%\collaboration{}
%\noaffiliation

\date{\today}

\begin{abstract}
Agglomeration processes occur in many different realms of science such as colloid and aerosol formation or
formation of bacterial colonies. We study the influence of primary particle density in agglomerate structure
using diffusion-controlled Monte Carlo simulations with realistic space scales through different regimes (DLA
and DLCA). The equivalence of Monte Carlo time steps to real time scales is given by Hirsch's hydrodynamical
theory of Brownian motion. Agglomerate behavior at different time stages of the simulations suggests that three
indices (fractal exponent, coordination number and eccentricity index) characterize agglomerate geometry. Using
these indices, we have found that the initial density of primary particles greatly influences the final
structure of the agglomerate as observed in recent experimental works. % \cite{Chak1}.
\begin{description}
\item[PACS numbers]{82.70.Rr, 61.43.Hv, 02.70.Uu, 05.10.Ln}
\item[Keywords]{Brownian motion, Monte Carlo methods, fractal-like aggregates.}
\end{description}
\end{abstract}

% insert suggested PACS numbers in braces on next line
%\pacs{82.70.-y, 83.10.-y, 61.43.Hv, 02.70.-c, 05.40.-a}
% insert suggested keywords - APS authors don't need to do this
%\keywords{Monte Carlo}

%\maketitle must follow title, authors, abstract, \pacs, and \keywords
\maketitle

% body of paper here - Use proper section commands
% References should be done using the \cite, \ref, and \label commands
\section{INTRODUCTION}
\hspace*{.3cm} Agglomeration of single particles to generate larger aggregates is an ubiquitous physical
phenomenon in nature. Not only the physics and chemistry of colloids and aerosols is governed by agglomeration
but also more complex mechanisms occurring in proteins or viruses depend on it \cite{Marc}. Self-propelling
{\em active} particles such as bacteria, insects, birds or fish may agglomerate to form colonies swarms, flocks
or schools \cite{Marc}. Agglomeration of active or passive particles involving Brownian motion is quite common.
Inert particles arising from combustion processes may aggregate forming aerosols and soot agglomerates. Soot
agglomeration is very important for industry and everyday life. Particulate matter generated during combustion
may have undesired effects including corrosion of boiler surfaces caused by particle deposition and chemical
activity (\textit{fouling}), deposition, chemical activity and particle fusion (\textit{slagging}) \cite{FCC},
and serious health problems such as pneumoconiosis and lung cancer \cite{Hidy}.\\
\hspace*{.3cm} Aliphatic and aromatic compounds of hydrocarbons (present in tars) volatilize very quickly (with a characteristic
time of about $10^{-4}$ sec. \cite{mal}) and undergo subsequent chemical reactions, leading to the formation of
soot particles (which have lost most of their original hydrogen) and polyaromatic hydrocarbons. Primary
soot particles are mainly aggregates of thousands of graphitic crystallites whose size is about tens of nanometers
\cite{CGC}. These aggregates tend to stick together immediately after their formation forming ``fractal-like"
structures. This is the agglomeration process. Additional processes like sintering will affect the shape and
properties of the agglomerates at much longer times. Here we want to describe the whole agglomeration process
since its early stages and studying the evolution of agglomerate structure. This is important e.g. for
understanding vapor condensation on agglomerates in boundary layer flows near the walls of a combustion chamber.\\
\hspace*{.3cm} The literature contains numerous models of agglomeration processes. Many of these works use Langevin equations
\cite{Mou,Dros} or kinetic (Smoluchowski) equations written in terms of a collision frequency factor.
Many others resort to numerical simulations based on some random-like collision algorithm, e.g. Monte Carlo
simulations, based on the solution of the Langevin equation in integral form. A detailed historical review is
\cite{mea99}. Most importantly, the agglomerates obtained through all these models are fractal-like structures,
a consequence that has been validated by many experimental works \cite{Sam,Sor97,Sor00,Mul}. Many works simulate
agglomeration starting from a number of particles in a given volume  \cite{Mou,Dros,mea85,par01,cho11}. In
these works, the calculated fractal exponents of  agglomerates range from 1.62 to 1.9, the particle number
density is between $10^{14}$ and $10^{15}$ cm$^{-3}$ whereas the expected amount of soot in a combustion chamber
is in the range of much lower values, $10^{10}\sim 10^{12}$ cm$^{-3}$ \cite{Sam}.
In these works, numerical simulations yield fractal exponents of agglomerates between 1.7 and 1.8 which correspond
to three-dimensional diffusion-limited colloid aggregation (DLCA). Recently Chakrabarty et al. \cite{Chak1} have
observed soot fractal aggregates with fractal exponents in the range $1.2\sim 1.5$ from ethene-oxygen premixed
flames with $2.3\sim 3.5$ fuel-to-air equivalence ratio. These exponents are noticeably lower than DLCA values
(about 1.8 \cite{mar07}).\\
Although Langevin equations seemingly provide very appropriate ways to tackle agglomeration, their use has been
marred by different shortcomings. For instance, Isella and Drossinos \cite{Dros} write a Langevin equation for
each monomer which is computationally quite costly. Moreover, the equivalent physical time of their simulations
is very short because the agglomerates dissolve quickly after being formed. They also use an extremely high
particle density, evidently to accelerate the agglomeration process. Mountain et al. \cite{Mou} save computation
 time by the crude simplification of considering a generic Langevin equation for noninteracting particles. \\
Here we reproduce the agglomeration process through a Monte Carlo simulation considering that both single particles
and the resulting agglomerates undergo Brownian motion. Brownian motion decreases as the particles collide and
bond, i.e. agglomerates move more slowly than particles. This speed reduction is due to the increasing frictional
resistance of the carrier gas. Different ways of incorporating the frictional resistance of the carrier gas into
the simulations include using empirical expressions of diffusive mobility in fractal aggregates coming from
laboratory measurements \cite{WanS} and assimilating the agglomerate to a porous medium \cite{TR95,TR96}. A
common feature of these methods is that in some step of the process the agglomerate is characterized by a single
parameter, which misses somewhat the agglomerate geometry. Instead, we have used the Riseman-Kirkwood theory
that incorporates the geometrical configuration of the whole agglomerate in the calculation of the diffusivity
(which is obtained at each time step of the simulation)\cite{Rheo}. Starting from a uniform spatial distribution,
particles and agglomerates move randomly move and interact in a 3D cubic lattice. They evolve from an initial
stage of diffusion limited aggregation (DLA), in which clusters grow by aggregating one single particle at a time \cite{halsey}, to a later DLCA stage, in which clusters stick to clusters. In order to
obtain the equivalency of Monte Carlo times and physical times, the time elapsed during a simulated Brownian
jump is calculated using Hinch's theory of Brownian motion \cite{Hin75} which, being local, is consistent with
the Riseman-Kirkwood theory (See appendix). \\
In order to compare our work to previous experimental results \cite{Chak1}, our simulations have been run up to
6 seconds of equivalent physical time. We have found that the fractal exponents of the agglomerates increase
with particle density and we have also studied the effect of the latter on the evolution of the fractal exponents.
Another important point in our work is the geometrical characterization of the agglomerates. There are abundant
references in the literature to the influence of the prefactor and the fractal exponent in the morphology of the
agglomerate. Much more sparse are the references to the coordination number and its influence \cite{Dros}. In
addition to fractal exponent and coordination number, we introduce here the eccentricity index (which has some
precedent in the triangle distribution function \cite{Hutt}). The geometric mean of the two last indices shows
an unexpected regularity for different times, densities and morphologies.\\
The rest of the paper is organized as follows. In section 2, we describe the simulation algorithm. We describe
and discuss our results, validate the model and include a geometrical description of agglomerates in section 3.
Section 4 contains our conclusions and the Appendix is devoted to technical matters.\\

\section{Simulation details}
\begin{figure}[b]
\includegraphics[scale=.2]{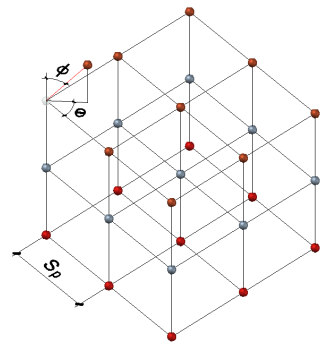}
\caption{\label{grid_} The cubic lattice used in the simulations. Spacing between particles, $S_{p}$ is equal
to $\displaystyle{\frac{1}{d_{p}}\left(\frac{1}{n_{p}}\right)^{1/3}}$,
where $n_{p}$ is the primary particle number density.
Polar and azimuth angles are taken respecting to a coordinate system that moves with the particle or the
centroid of the agglomerate.}
\end{figure}

\hspace*{.3cm} Initially, $N_{p}=8000$ particles of diameter $d_p=50$ nm occupy the nodes of a cubic
lattice of side $S_p$ \cite{PA} (see Figure \ref{grid_}). Then the particle number densities are in the range
$10^{10}\,\sim\,10^{14}$ cm$^{-3}$. Notice that expected soot particle densities inside combustion chambers are
 in the order of $10^{10}\,\sim\,10^{12}$ cm$^{-3}$ \cite{Mor}. As we explain in the next section, the selected
number of particles allows the particle distribution function to become self-similar, with quasi-steady moments,
thereby avoiding boundary effects. Self-similar size distributions have been widely observed in aerosols
\cite{fri}.

In the simulations, particles undergo Brownian motion with a fixed length step (which is $h_p=2d_p$ for single
particles) but they move in random directions given by the angles shown in Figure \ref{grid_}. The relation
between the simulation time step and real time is such that the root mean square displacement of the particle
during a time step is $2d_p$. Then \cite{berne} (Appendix 5A), \cite{wei89}
\begin{equation}
\sqrt{\langle\Delta x(t)^2\rangle}=2d_p,\, \langle\Delta x(t)^2\rangle=2\int_0^t(s-t)\langle v_x(0)v_x(s)
\rangle ds.\label{time_step}
\end{equation}
In the simplest case, the velocity autocorrelation is that of an Ornstein-Uhlenbeck process \cite{Chand,Capas},
whereas a more realistic expression is provided by Hinch's theory of Brownian motion \cite{Hin75}. Both mean
squared displacements are listed in the Appendix. Inserting these expressions in (\ref{time_step}) and solving
that equation for the time step, we obtain Ornstein-Uhlenbeck and Hinch times. The largest of the two is usually
the Hinch time which we select as our time-step. The random polar ($\Phi=\pi\,\delta_{1}$) and azimuthal
($\Theta=2\,\pi\,\delta_{2}$) angles are referred to a coordinate system that moves with each particle.
$\delta_{1}$ and $\delta_{2}$ are random numbers uniformly distributed between 0 and 1. This choice avoids
unequal chance fluctuations in long sequences of random angles \cite{Fell1} and it does not require to randomly
generate angles out of a uniform spherical distribution as in \cite{Wozn}. We have used periodic boundary
conditions to preserve the particle density during the simulation, i.e. particles that move out of the domain
are re-injected from the opposite boundary.\\
As agglomeration criterion, we consider that two particles (single or pertaining to an agglomerate) whose
centers get closer than $2\,d_p$ will agglomerate. We prevent overlapping by calculating if the jump length
leading to collision along a random direction is smaller than $2\,d_{p}$. As the agglomerates increase their
size, drag forces due to friction with the carrier gas increase too and we expect a reduction in the agglomerate
velocity. To take into account this effect, we calculate the translational diffusion coefficient of each
agglomerate, anytime a new bond is formed, by means of the Riseman-Kirkwood theory \cite{Rheo}:
\begin{eqnarray}
D_{a}&=&\displaystyle{\frac{D_{p}}{N_{a}}\left(1+\frac{d_{p}}{2\,N_{a}}\sum\limits_{i}\sum\limits_{j}
\frac{1}{R_{ij}}\right)}\nonumber
\end{eqnarray}
where $D_{p}$ is the particle diffusion coefficient (see Appendix), $D_{a}$ is the agglomerate diffusion
coefficient, $N_a$ is the number of
particles in the agglomerate and $R_{ij}$ is the distance from $i^{th}$ to $j^{th}$ particles in the agglomerate.
Notice that as the size of the agglomerate increases, $N_a$ increases and then, the diffusion coefficient tends
to decrease. The Brownian jump of each agglomerate, $h_a$, is given by the following simple rule:
\begin{eqnarray}
h_{a}=h_p \frac{D_a}{D_p}\nonumber
\end{eqnarray}
A jump length of $2\,d_{p}$ only occurs for primary particles: the agglomerates jump over smaller distances as
they grow. Our algorithm is summarized in the following lines:\\

\begin{enumerate}
\item Set the number of particles and distribute them homogeneously in a cubic lattice.
\item Choose a particle number density which leads to a particle spacing ($S_{p}$) for initial distribution.
\item Fix the size of the primary particle jump ($h_p=2d_p$) and calculate the physical time ($t_{step}$)
corresponding to one simulation time step.
\item Pick a maximum real time for the simulation ($T$)
\item While t $<T$
\begin{enumerate}
\item Generate random angles for each particle/agglomerate in the simulation.
\item Produce jumps and update positions.
\item Check distances between external particles and agglomerates and join to the latter all the external
particles within a distance $<2d_p$ (to prevent overlapping).
\item Update diffusion coefficients and jump lengths ($h_a$).
\item t = t + $t_{step}$.
\end{enumerate}
\item End
\end{enumerate}

Most Monte Carlo simulations of agglomeration processes use different algorithms for the DLA and DLCA stages \cite{mea99,mea85,cho11}. The particles or clusters that join together and the way they join are determined by a random {\em ad hoc} procedure that does not have a correspondence to the actual physical system. Instead, we use a single algorithm that does not distinguish between DLA and DLCA stages. In our algorithm, contacts between particles and clusters and clusters with clusters occur as a result of the Brownian motion of particles and clusters themselves in a real-scale space. We have ignored cluster rotation by simplicity.
%A diferencia de una gran cantidad de simulaciones tipo Monte Carlo de procesos de aglomeracion \cite{mea99,mea85,cho11}, que utilizan algoritmos distintos para las etapas DLA y DLCA y donde las particulas o los clusters que se unen, asi como el modo en que lo hacen, se determinan mediante un procedimiento aleatorio establecido \textit{ad hoc} sin un equivalente fisico real, nosotros desarrollamos un algoritmo unico donde los contactos particula-cluster-cluster se producen como resultado de los movimientos estocasticos (que reproducen el movimiento Browniano) de las particulas y los clusters en un espacio a escala real. Por simplicidad no se ha considerado el giro de los clusters / C-R1; E-R2}

\section{Numerical results and discussion}

\hspace*{0.3cm}In our simulations, we have used parameter values corresponding to soot formation inside a
combustion chamber as indicated in Table 1. In the Table, $\rho_{\mathrm{p}}$, $d_p$, $n_p$, $T$, $P_a$,
$\mu_a$, $\rho_{\mathrm{a}}$ and $k_{B}$ are the soot density, the primary particle diameter, the primary
particle number density, the air temperature, the air pressure, the air viscosity (calculated using Sutherland
relation), the air
density (considered as an ideal gas) and the Boltzmann constant, respectively.\\

\begin{table*}
\begin{center}
\begin{tabular}{@{} cccccccc@{}}
{$\rho_{p}$} & {$d_p$} & {$n_{p}$} & {$T$} & {$P_{a}$} & {$\mu_{a}$} & {$\rho_{a}$} & {$k_B$} \\ [3pt]
(g cm$^{-3}$) & (nm) & (cm$^{-3}$) & (K) & (Pa) & (N s m$^{-2}$) & (kg m$^{-3}$) & (kg m$^2$ s$^{-2}$) \\
& & & & & & &\\
$2$ & $50$ &
$10^{10}\,\sim\,10^{14}$ & $1900$ & $1.01325\times10^{5}$ & $6.20\times10^{-5}$ & 0.173 & $1.38\times10^{-23}$
\\
\end{tabular}
\end{center}
\caption{Constants and parameters used in the simulation}
\label{table1}
\end{table*}

\begin{figure}[h!]
\includegraphics[width=9cm,height=6.25cm]{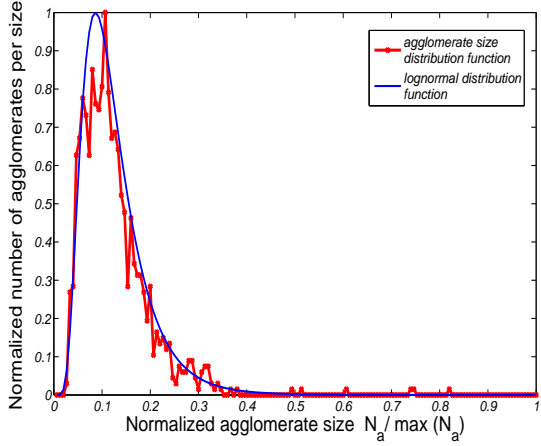}
\caption{\textit{Size distribution function in terms of the scaled size $\xi=i/N(t)$, where $N(t)$ is the
number of aggregates at time $t$. The simulation lasts 6 seconds and the initial particle density is $4\times
10^{10}$ cm$^{-3}$.}}
\label{lognormal_sp_60_6s_}
\end{figure}

We have computed 100 sets of simulations for a primary particle number density of $4\times10^{10}$ cm$^{-3}$
and 10 sets for other 9 different densities between
$10^{10}$ and $10^{14}$ cm$^{-3}$ in order to observe the effect of the primary particles density in the
resulting structures. 61 samples per simulation were taken to
study in detail the time evolution of the system. Simulations for the different densities were run up to
6 seconds of physical equivalent time. We compare the size distribution obtained
through our simulations with a lognormal distribution because the latter describes very well atmospheric aerosols, mainly those
coming from a single source \cite{Hinds}. This is shown in Figure \ref{lognormal_sp_60_6s_}
for a primary particle number density of $4\times10^{10}$ cm$^{-3}$.
For $N_{p}=8000$ particles, the size distribution function becomes self-similar after sufficient time. This implies
that there are no boundary effects. To further check this, we have calculated the time evolution of the geometrical
as well as the logarithmic moments (from the $2^{nd}$ to the $6^{th}$ moment),

\begin{eqnarray}
\displaystyle{\langle\eta^{k}\rangle}&=&\displaystyle{\frac{1}{N}\sum_i\eta_i^kn_i=\frac{\mu_{k}}{\mu_1^{k}}\,
N^{k-1}},\nonumber\\[1ex]%\quad\text{$k^{th}$ geometric moment of $f(\eta,t)$}\nonumber\\[1ex]
\displaystyle{\langle\left(\ln\eta\right)^{k}\rangle}&=&\displaystyle{\frac{1}{N}\sum\limits_{i}[\ln\eta_{i}]
^{k}\,
n_{i}},\nonumber\\%\quad\text{$k^{th}$ logarithmic moment of $f(\eta,t)$}\nonumber\\
\displaystyle{\mu_k}&=&\sum\limits_{i} i^k n_i, \nonumber
\end{eqnarray}
where $\displaystyle{\langle\eta^{k}\rangle}$ is the $k^{th}$ geometric moment of $f(\eta,t)$,
$\displaystyle{\langle\left(\ln\eta\right)^{k}\rangle}$ is the $k^{th}$ logarithmic moment of $f(\eta,t)$,
$n_i(t)$ is the number of agglomerates with $i$ particles at time $t$, $\eta_{i}(t)=\displaystyle{\frac
{iN(t)}{\mu_1(t)}}$ and $N=N(t)=\mu_0(t)$ is the number of agglomerates at time $t$. Assuming that the size
distribution function is self-similar, $n_i(t)=f(i/N(t))$, and we get
\begin{widetext}
\begin{equation}
\mu_k=\int i^kf\!\left(\frac{i}{N(t)}\right)\! di=[N(t)]^{k+1}\int \xi^k f(\xi)d\xi\Longrightarrow \langle
\eta^k\rangle=\frac{\int\xi^k f(\xi)d\xi}{[\int\xi f(\xi)d\xi]^k},\nonumber
\end{equation}
\end{widetext}
so that the $k^{th}$ moment is independent of time. Similarly, the logarithmic moments should be independent
of time once the self-similar size distribution is established.

\begin{figure}[h!]
\includegraphics[width=8.5cm]{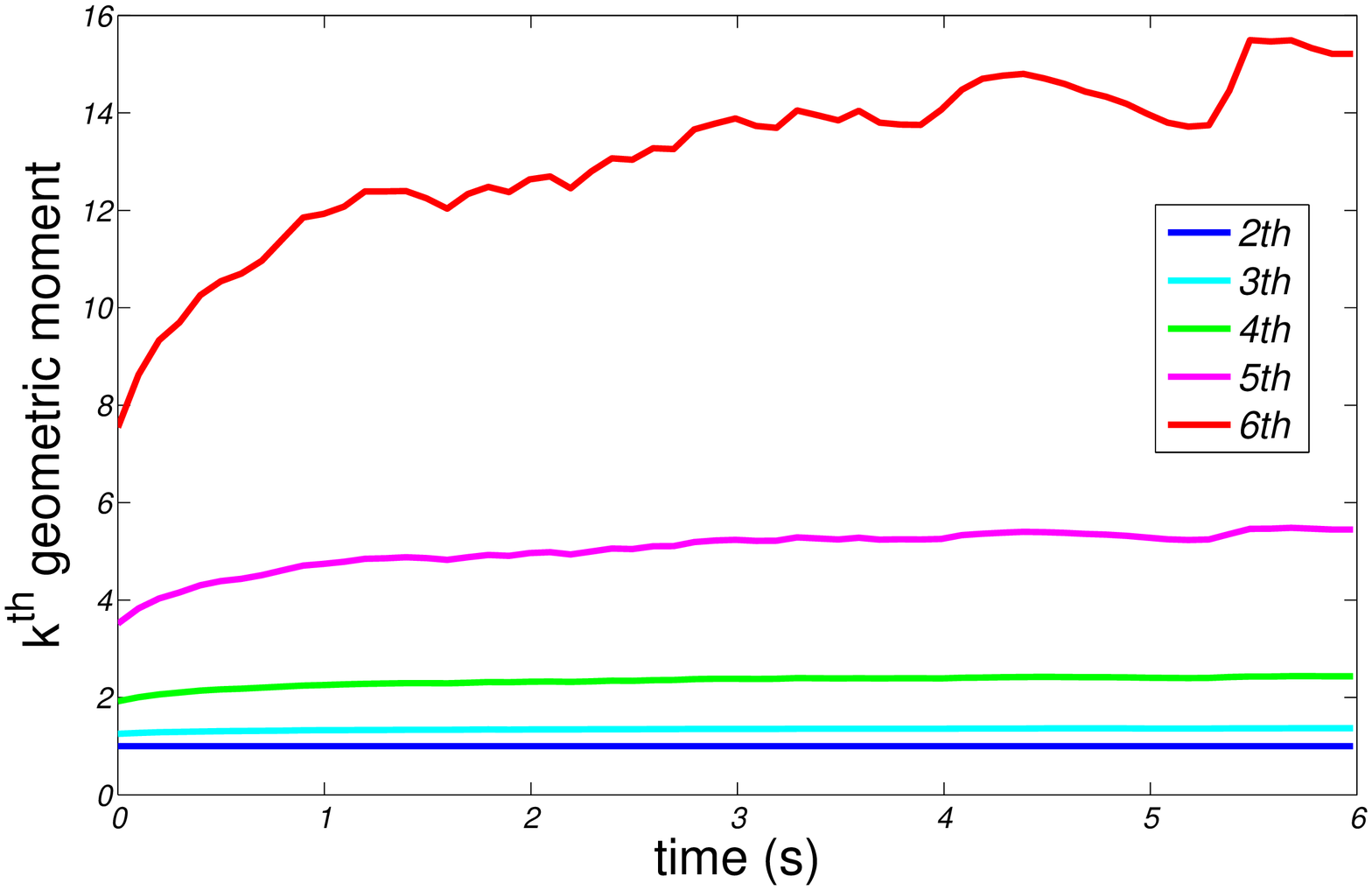}
\includegraphics[width=8.5cm]{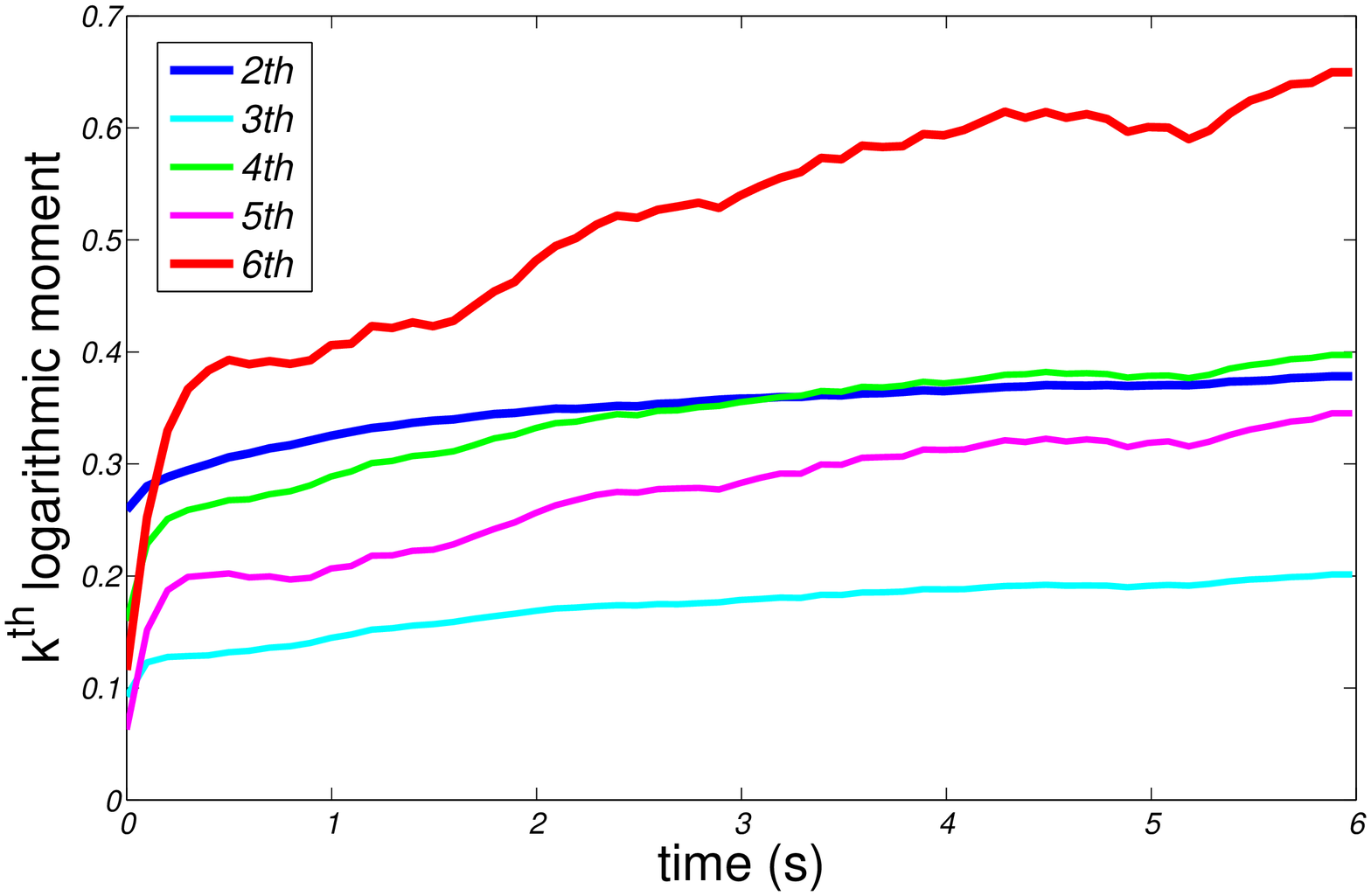}
\caption{\label{moment} \textit{Time evolution of the averaged geometric (left) and logarithmic (right)
$\mathit{k^{th}}$ moments, obtained from the simulations.}}
\end{figure}

Figure \ref{moment} shows the averaged geometric (left) and logarithmic (right) moments for a total number of
$8000$ particles. For a physical time of 6 seconds, the lower order geometric and logarithmic moments reach a
steady state which indicates that a self-similar size distribution function has been reached. This indicates
that we obtain reliable results from simulations with 8000 particles.

\begin{figure}[h!]
\includegraphics[width=8cm,height=5cm]{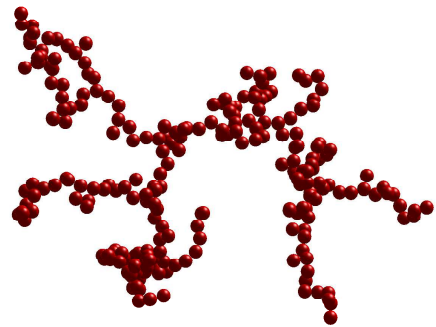}\\
(a)\\[2ex]
\includegraphics[width=8cm,height=5cm]{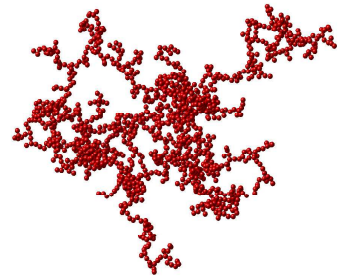}\\
(b)\\[2ex]
\includegraphics[width=8cm,height=5cm]{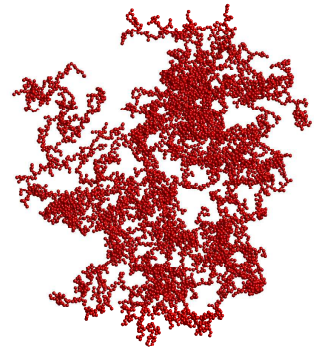}\\
(c)
\caption{\textit{Different sized agglomerates: (a) with 245 particles, (b) with 1484 particles and (c) with
8000 particles, corresponding to
densities $4\times10^{10}$, $10^{12}$ and $10^{14}$ cm$^{-3}$, respectively.}}
\label{245_sp_60}
\end{figure}

\begin{figure}[h!]
\includegraphics[width=9cm,height=6cm]{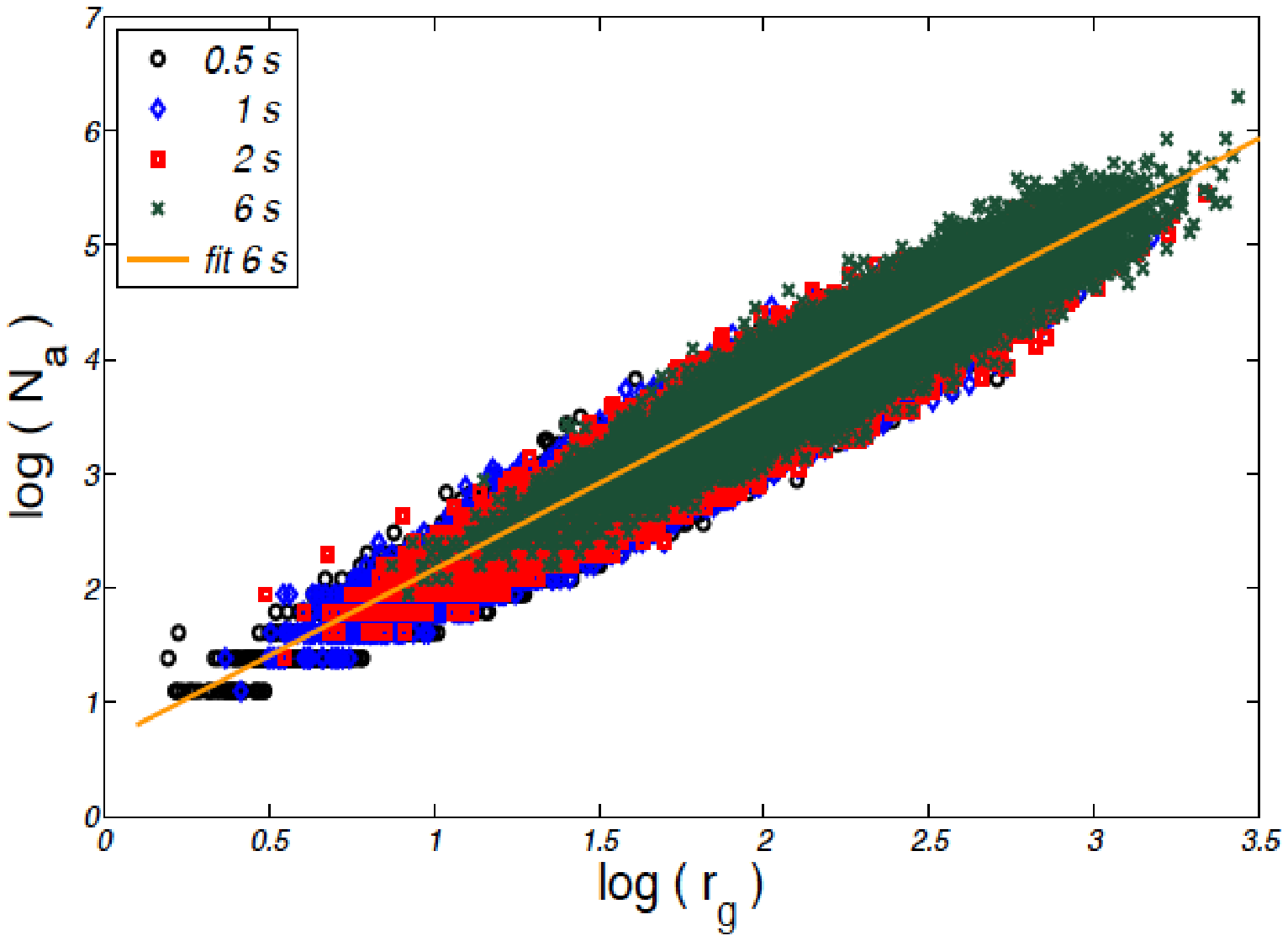}
\caption{\label{log_fit} \textit{Number of particles per agglomerate versus radius of gyration for a primary particle density of $4\times10^{10}$ cm$^{-3}$ and equivalent physical times of 0.5, 1, 2 and 6 s. Different points correspond to different simulations and we have fit a straight line through the points corresponding to 6 s. Note that there is no indication of transient behavior}.}
\end{figure}

\subsection{Fractal exponents of the agglomerates}
\begin{figure*}
\includegraphics[width=18cm,height=10cm]{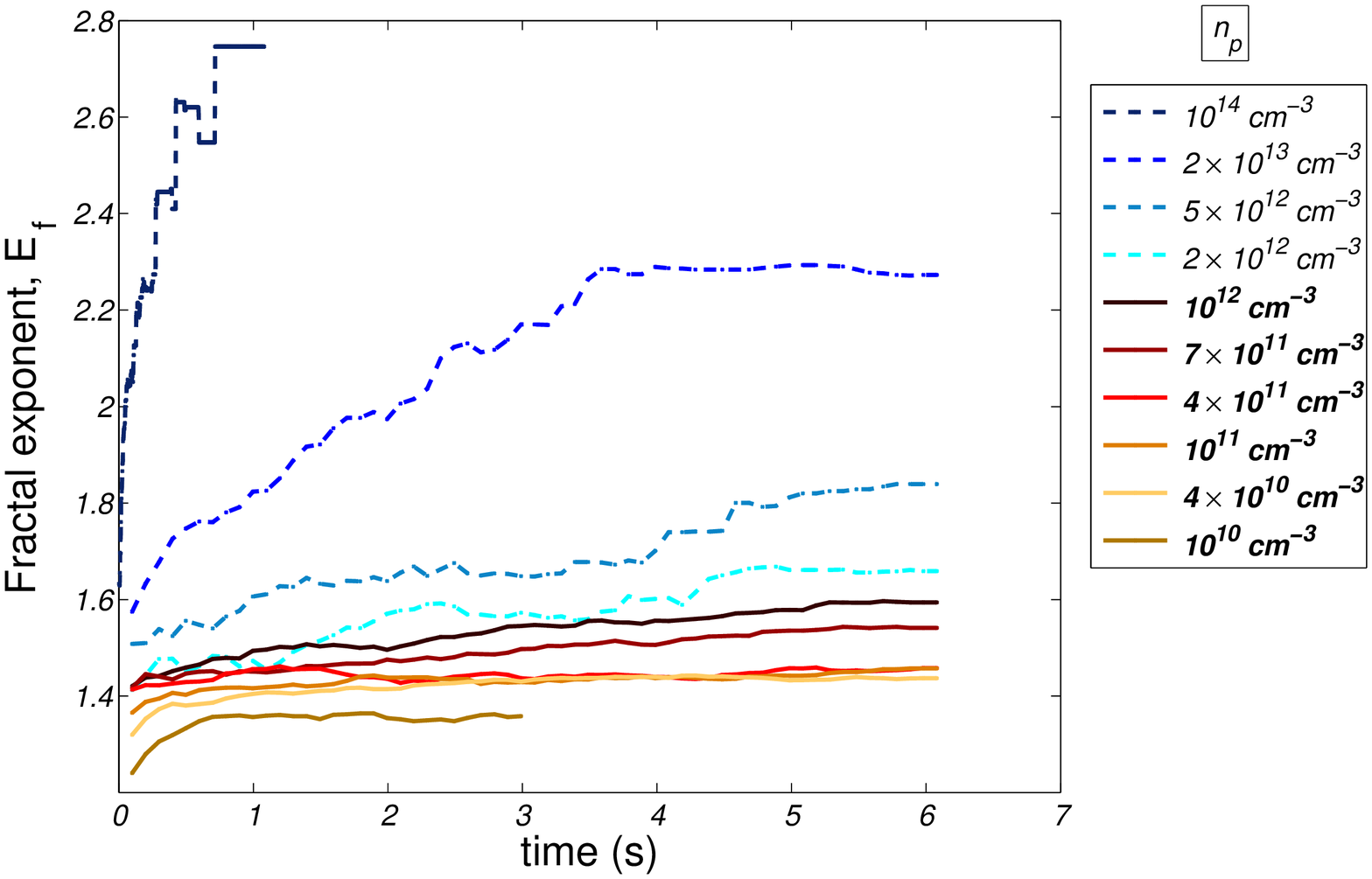}%{Ef_t__.eps}
\caption{\label{evolution_density_last} \textit{Time evolution of the fractal exponent $E_{f}$ for different primary particle
(soot) number densities $n_p$. The exponents corresponding to densities seen in combustion processes are
depicted as solid lines. The uppermost curve is interrupted because an agglomerate comprising all the
particles is formed before 6 seconds.}}
\end{figure*}

\begin{figure}[h!]
\includegraphics[width=12cm,height=6cm]{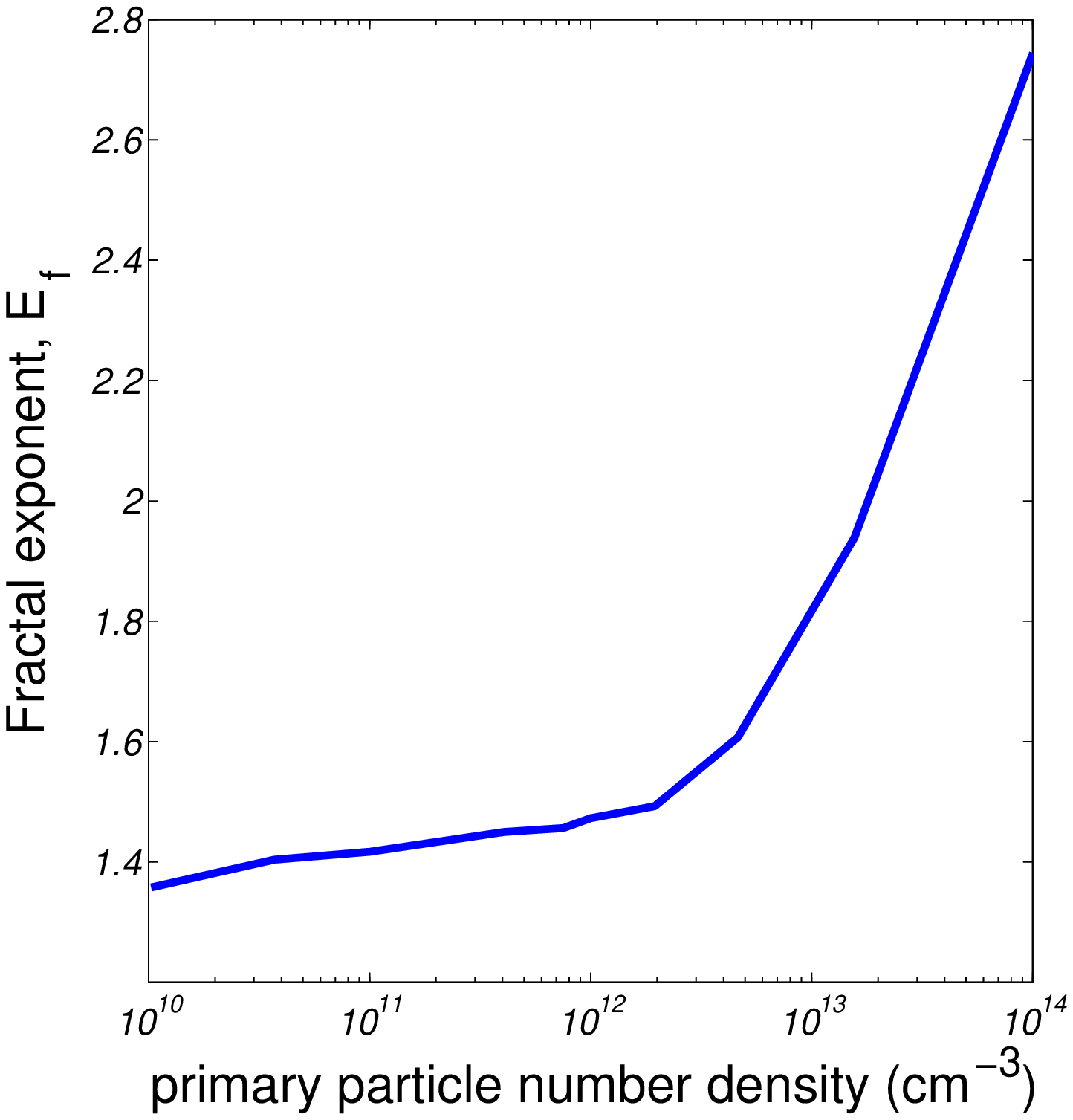}
\caption{ \textit{Fractal exponent $E_{f}$ versus primary particle number density $n_p$
after 1 second. The steeper part, at the right, is indicating that bigger clusters are formed more rapidly
due to the higher densities of primary particles} ($>2\times10^{12}$ cm$^{-3}$). }\label{Ef_rho_t_1s}
\end{figure}

Figure \ref{245_sp_60} shows three agglomerates with 245, 1484 and 8000 particles for primary particle number
densities of $4\times10^{10}$, $10^{12}$ and $10^{14}$ cm$^{-3}$, respectively. It can be appreciated that very
open fractal-like structures appears for low density values, evolving to more compact shapes as the density
increases. In an agglomerate, the number of particles $N_a$ is related to the radius of gyration $r_{g}$ (mean
squared radius) by $N_a=k_a r_g^{E_f}$, where $E_f$ is the mean fractal exponent $E_{f}$ and $k_a$ is a
prefactor \cite{fri}. From the linear fit,
\begin{eqnarray}
&&\log(N_a) = \log(k_{a}) + E_f \, \log(r_{g}), \nonumber\\[1ex]
&&r_g^2=\frac{1}{N_a}\sum_{j=1}^{N_a}(\mathbf{r}_j-\mathbf
{\bar{r}})^2=\frac{1}{2N_a^2}\sum_{i,j=1}^{N_a}(\mathbf{r}_i-\mathbf{r}_j)^2\nonumber
\end{eqnarray}
(where $\mathbf{r}_j$ is the position of particle $j$ in the agglomerate of size $N_a$ and $\mathbf{\bar{r}}$
is the mean position), we can extract the fractal exponent $E_f$ that characterizes agglomerates; see
Figure \ref{log_fit}. Figure \ref{evolution_density_last} shows $E_f$ varies with time for different values of the primary particle number density (from
$10^{10}$ cm$^{-3}$ to $10^{14}$ cm$^{-3}$). After a 1 second equivalent physical time, the maximum $E_f$
varies between 1.4 and 2.8 for the considered densities, as depicted in Figure \ref{Ef_rho_t_1s}. The fractal exponent tends to a constant and larger value for larger times (with one exception). Thus the fractal exponent obtained after one second is a lower bound of the asymptotic value of the fractal exponent. In the case of the outlier, with primary particle number density of $10^{14}$ cm$^{-3}$, one second is sufficient for bringing to completion the agglomeration process. To attain the asymptotic value of the fractal exponent in that case, we should have used a much larger value of the total number of particles which would have increased considerably the computational cost. The average agglomerates fractal exponents reach
values between 1.4 and 2.8 for the range of initial particle densities we use.

In the literature, the calculated fractal exponents of  agglomerates range from 1.62 to 1.9, for particle
number densities between $10^{14}$ and $10^{15}$ cm$^{-3}$ \cite{Mou,Dros,mea85,par01,cho11}. These fractal
exponents are in the range of 3D DLCA, about 1.8 \cite{mar07}.
Recently Chakrabarty et al.\ have observed soot fractal aggregates with much lower fractal exponents in the
range $1.2\sim 1.5$ from ethene-oxygen premixed flames with $2.3\sim 3.5$ fuel-to-air equivalence ratio. While
these fractal exponents are lower than those found in the literature \cite{Mou,Dros,mea85,par01,cho11}, the
initial number density in the experiments is in the range $10^{10}\sim 10^{12}$ cm$^{-3}$, which is also lower
than the $n_p$ values used in the numerical works. The fractal exponents observed in experiments are like those
found in our simulations for the same number density range. Note that the fractal exponent rises more abruptly
for $n_p$ above $10^{12}$ cm$^{-3}$ to within the DLCA range found in \cite{Mou,Dros,mea85,par01,cho11}.

\subsection{Geometrical characterization of the agglomerates}
In addition to the fractal exponent, we may characterize agglomerates by other indices of geometrical nature.
In an agglomerate, the relative number of particles $n_g^j$ surrounding a given one $j$ (at a distance not
larger than $2d_p$) gives an idea of the compactness of the latter, and we call it coordination index,
$$i_{c}^{j}=\displaystyle{\frac{n_{g}^{j}}{12}}\in [0,1],\qquad\text{coordination index of $j^{th}$ particle.}
$$
$i_c^j=0$ corresponds to an isolated particle and $i_c^j=1$ gives close packing of the particle. The coordination
number defined as in \cite{Dros} is twelve times our coordination index.

We have also defined the eccentricity index as follows:
\begin{eqnarray}
i_{e}^{j}&=&\displaystyle{\frac{|\mathbf{r}_{CM}-\mathbf{r}_j|}{r_{e}}},\qquad\text{eccentricity index of $j^{th}$
particle}\nonumber
\end{eqnarray}
where $\mathbf{r}_{CM}$ is the position of the center of mass of the system formed by the $j^{th}$ particle at
$\mathbf{r}_j$ and its surrounding neighbors (at distances no larger than $2d_p$), and $r_{e}$ is the {\em
enveloping radius} that corresponds to the maximum distance between the center of mass of the system and the
center of the neighbors surrounding the $j^{th}$ particle. This eccentricity index measures the way the
particles connect in an agglomerate. A particle with $i_{e}=0$ is surrounded in a spherically symmetric way,
whereas a particle with $i_{e}=1$ has the most asymmetric distribution of its surrounding particles.

\begin{figure}[b]
\includegraphics[width=9cm,height=5cm]{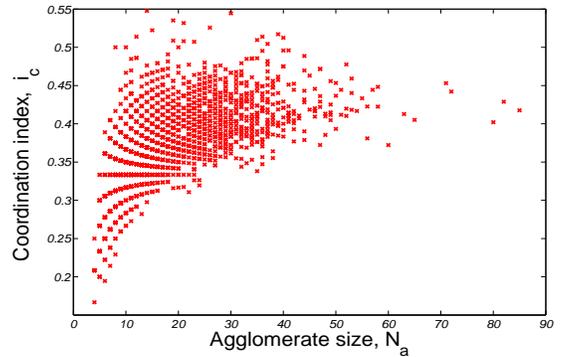}
\caption{\textit{Coordination index of aggregates in terms of its size for different simulations and a primary
particle number density $n_p=4\times 10^{10}$ cm$^{-3}$.}}
\label{arbo_sp_60_1s_}
\end{figure}

\begin{figure}[b]
\includegraphics[width=10cm,height=5cm]{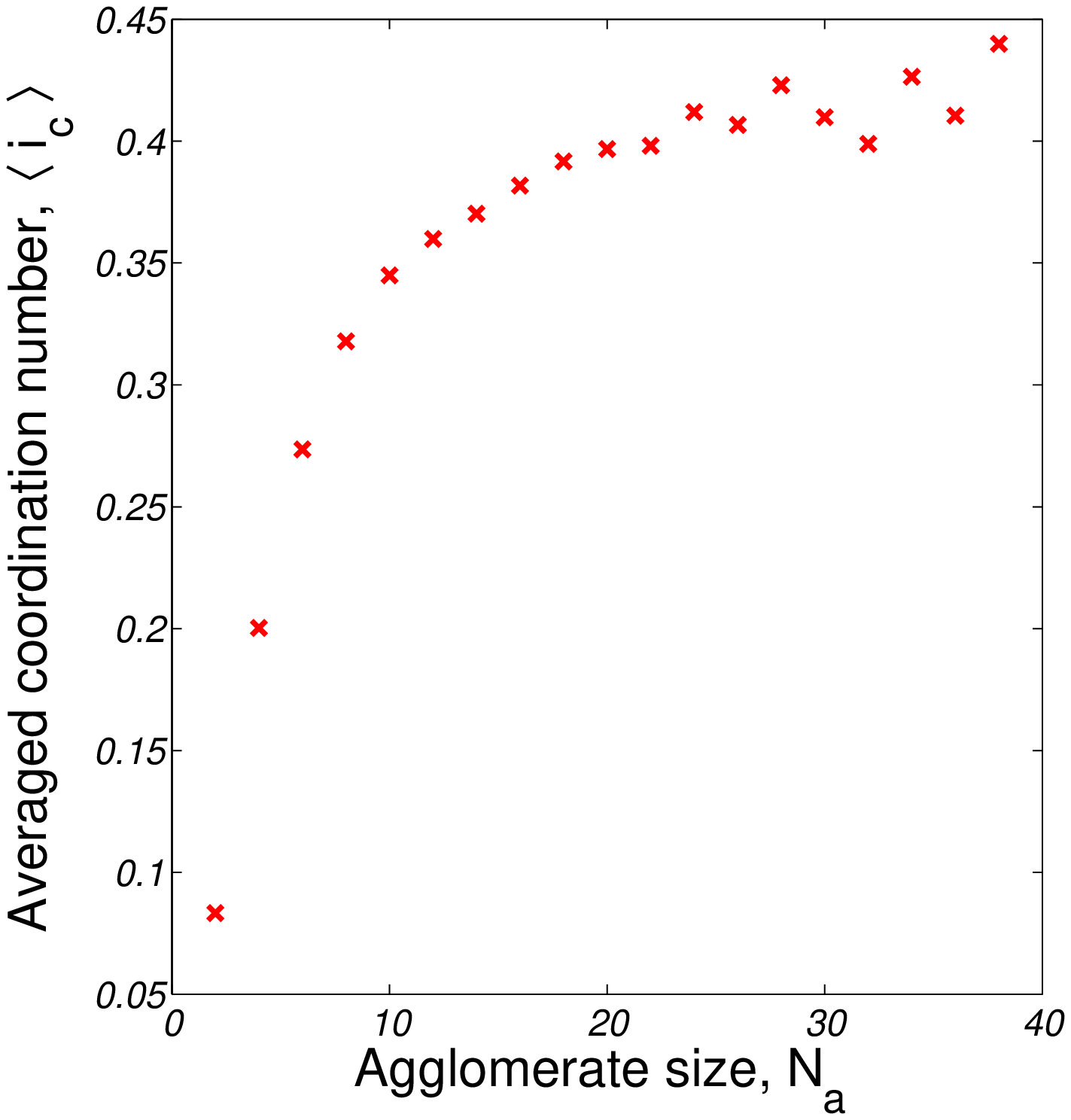}\\
(a)\\[2ex]
\includegraphics[width=10cm,height=5cm]{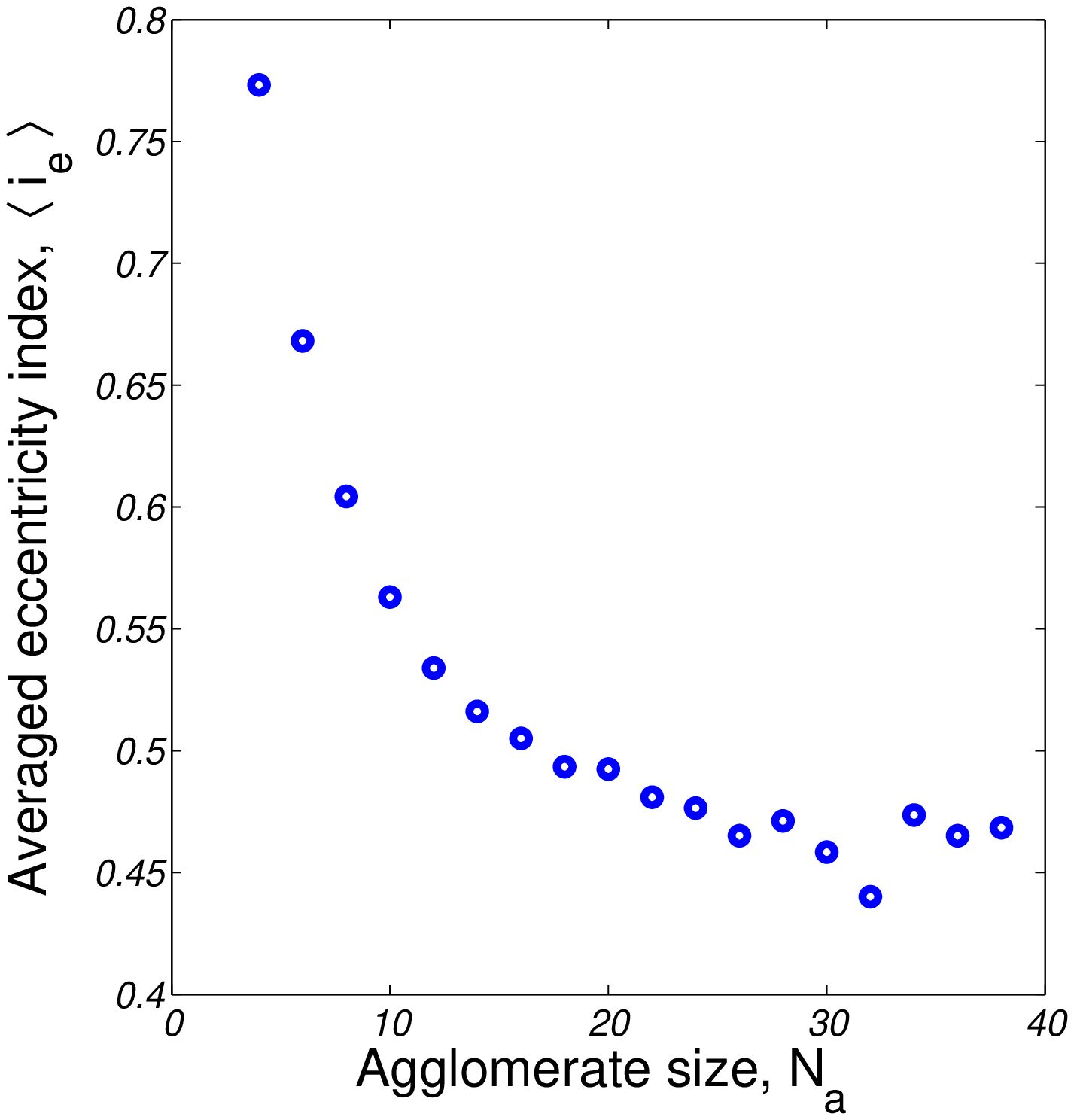}\\
(b)
\caption{\textit{(a) Variation of the average coordination index $\langle i_{c}\rangle$ with agglomerate size,
$N_a$. (b)Variation of the mean eccentricity index $\langle i_{e}\rangle$ with agglomerates size $N_{a}$.
Primary particle (soot) number density in both plots is $10^{10}$ cm$^{-3}$ for an equivalent physical of 1
sec}.}
\label{ic_Na_sp92_t_1s}
\end{figure}

\begin{figure}[h!]
\includegraphics[width=10cm,height=5cm]{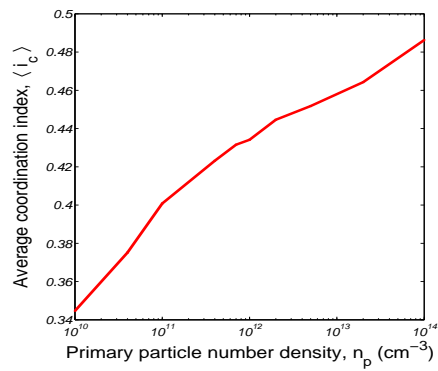}
\caption{\label{coord_vs_dens} Average coordination index $i_{c}$ of all aggregate sizes
versus primary particle number density $n_p$ after 1 second equivalent physical time.}
\end{figure}

\begin{figure}[b]
\includegraphics[width=6cm,height=4cm]{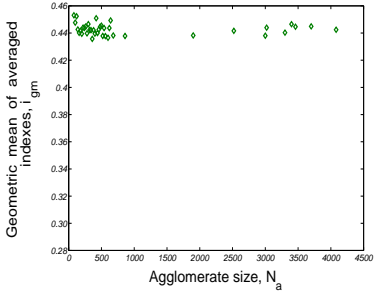}\\
(a)\\[2ex]
\includegraphics[width=6cm,height=4cm]{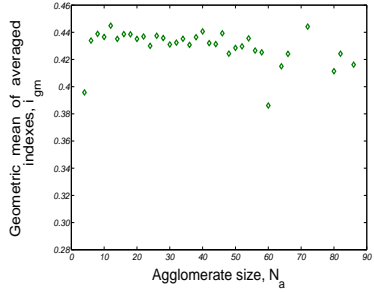}\\
(b)\\[2ex]
\includegraphics[width=6cm,height=4cm]{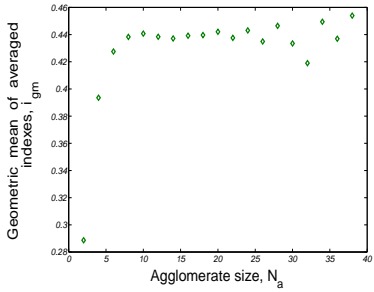}\\
(c)\\[2ex]
\caption{\textit{Geometric mean of the average coordination and eccentricity indices vs agglomerate size after
1 second for three different values of $n_p$: (a) $4\times 10^{10}$ cm$^{-3}$, (b) $10^{12}$ cm$^{-3}$ and (c)
 $10^{14}$ cm$^{-3}$.}}
\label{mean_mediag_sp_8_1s_}
\end{figure}

\begin{figure}[h!]
\includegraphics[width=8cm,height=5cm]{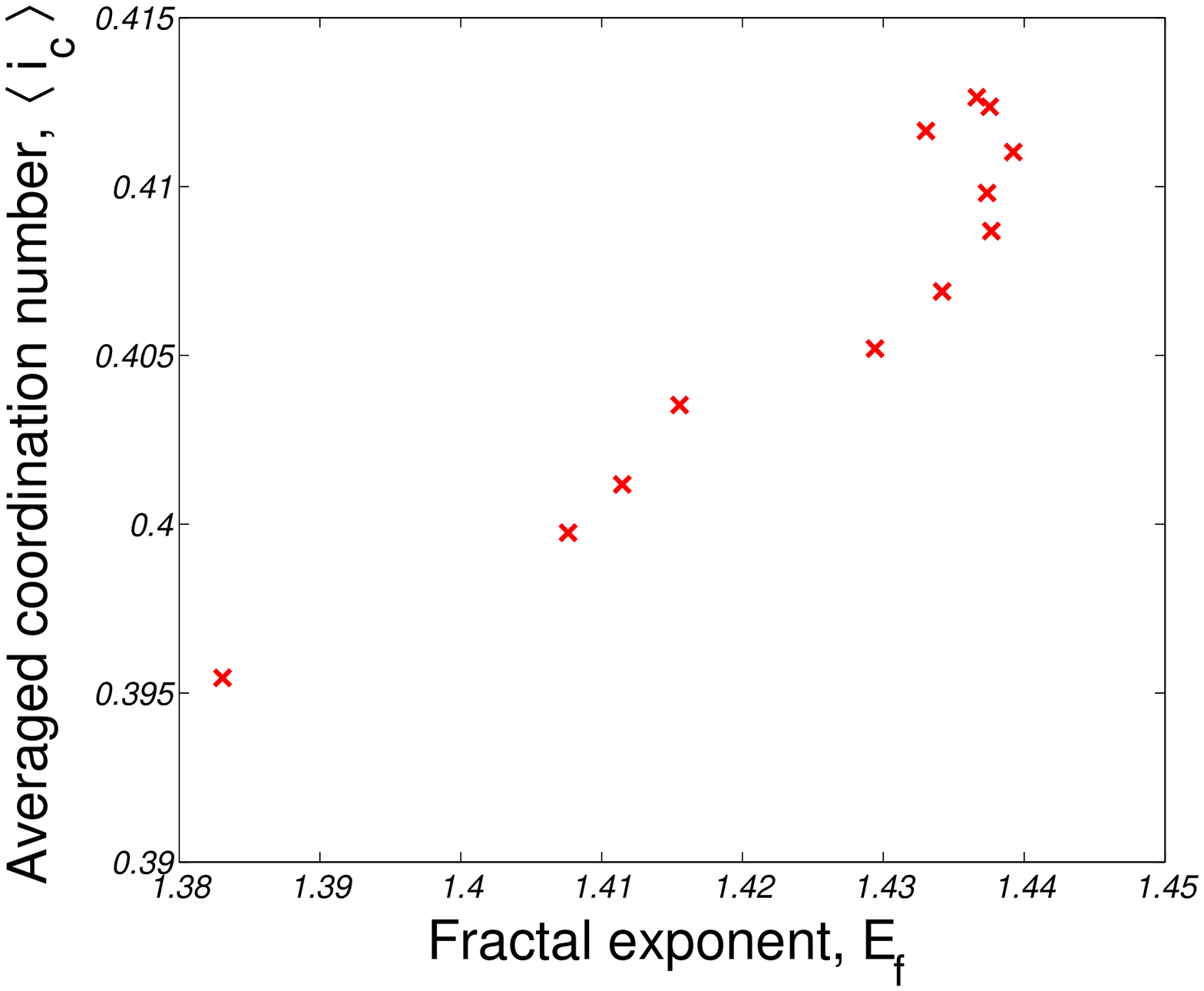}
\caption{\label{Ef_ic_sp60_} \textit{Evolution of the fractal exponent and the coordination index, for a primary particle density of} $4\times10^{10}$
\textit{cm}$^{-3}$, \textit{considering sizes} $>15$. Each point represents values at a given time and, as time increases, these points tend to accumulate in the upper part of the figure. }
\end{figure}

The coordination index of an agglomerate is calculated as the mean value of the coordination indices of all the
particles comprising it. The same applies for the eccentricity index. These coordination and eccentricity
indices depend on the agglomerate size $i$, the realisation of the Brownian motion $\omega$ and the particle
number density $n_p$, $i_{c,e}(i,n_p,\omega)$. For a given value of primary particle number density, the
coordination indices versus size for different realisations of noise (corresponding to different simulations)
are depicted in Figure \ref{arbo_sp_60_1s_}. The expected values of the indices (over all simulations) are the
average indices $\langle i_{c}\rangle(i,n_p)$ and $\langle i_{e}\rangle(i,n_p)$. Figures \ref{ic_Na_sp92_t_1s}
(a) and (b) show the average coordination and eccentricity indices for $n_p=10^{10}$ cm$^{-3}$. Note that
$\langle i_{c}\rangle$ increases with agglomerate size, whereas $\langle i_{e}\rangle$ decreases. As the
agglomerate size increases, the agglomerates change from being stringy structures with low $\langle i_{c}
\rangle$ and large $\langle i_{e}\rangle$ to becoming more compact, with both indices about 0.45; see Figure
\ref{245_sp_60}. Figure \ref{coord_vs_dens} shows the variation of the average coordination index of all the agglomerate sizes
(calculated after a 1 second equivalent physical time) with the primary particle number density. Similarly to the fractal
exponent behavior in Fig.~\ref{Ef_rho_t_1s}, this index increases with the density $n_p$.

The coordination and eccentricity indices and their geometric mean have the following properties:
\begin{itemize}
\item The plot of $i_{c}(i,n_p,\omega)$ as a function of $i$ in Figure \ref{arbo_sp_60_1s_}, for $n_p=4\times
10^{10}$ cm$^{-3}$ and different realisations, has a very organized pattern for low $n_p$ but does not present a recognisable
structure for high $n_p$ (e.g., for $10^{14}$ cm$^{-3}$).
\item $i_c$ and $i_e$ exhibits an asymptotic behavior for large $N_a$ as Figure \ref{ic_Na_sp92_t_1s} shows (see also \cite{WebFri,Dros}). The evolution to constant values of these geometric parameters and of the fractal exponent for large times is a sign that the internal structure of the agglomerate tends to become self-similar. The aggregates grow with time and, as they become larger, they become closer to self-similar and some connectivity pattern is repeated.
\item The average indices $\langle i_{c}\rangle$ and $\langle i_{e}\rangle$ probe the local structure of aggregates and they seem to be related for large aggregate size. Their geometric mean, $i_{gm}=\sqrt{\langle i_{c}\rangle\,\langle i_{e}\rangle}$, becomes almost constant for large $N_a$, as shown in Figure \ref{mean_mediag_sp_8_1s_} for three different densities. %$i_{gm}=\sqrt{\langle i_{c}\rangle\,\langle i_{e}\rangle}$ is always bounded in a range between 0.4 and 0.5 for all times and densities.
As the aggregates size $N_a$ grows, the increasing coordination index and the decreasing eccentricity index seem to compensate. Assuming an {\em ad hoc} very dense particle packing representing an upper limit for $i_{c}$, we have created a sequence of configurations and obtained the geometric mean $i_{gm}$ which is bounded between 0.4 and 0.5. %The almost constancy of the geometric mean (Figure \ref{mean_mediag_sp_8_1s_}) for large agglomerates with independence of their large scale morphology (given by the fractal exponent) looks reasonable, because coordination and eccentricity indices operate in a local scale.
\item The cluster distribution evolves to become self-similar and, at the same time, the fractal exponent and the coordination index evolve to constant values as shown in the highest part of Figure \ref{Ef_ic_sp60_}, where points accumulate. Both the fractal exponent (Fig.~\ref{Ef_rho_t_1s}) and the average coordination index (Fig.~\ref{coord_vs_dens}) of all the agglomerate sizes
    increase with the primary number density $n_p$. This seems reasonable as they both probe the self-similar structure of the clusters and are therefore related.

\end{itemize}

%In Figure \ref{Ef_ig_sp60_} we have plotted the parallel evolution of the fractal exponent and the
%geometric mean of average coordination and eccentricity indices, for a primary particle density of $4\times10^{10}$ cm$^{-3}$,
%considering sizes $>15$.

\section{Concluding remarks}

\hspace*{.3cm}We have simulated the agglomeration of single particles for different initial number densities
by a Monte Carlo method. The range of initial number densities covers the values expected for soot particles in
combustion processes and also higher values used by other authors in their simulations \cite{Mou, Dros}.
Initially, 8000 particles occupy a cubic domain with periodic boundary conditions (to preserve particle density).
This size produces a self-similar log-normal size distribution function after a short time with quasi-steady
moments. After an equivalent physical time of one second, a self-similar size distribution is reached. The fractal
exponent increases with primary particle number density, first slightly and, beyond $n_p=10^{12}$ cm$^{-3}$,
more abruptly. Below that density, the fractal exponent is no larger than 1.5 and it remains so no matter the
duration of the process. For such low densities, particle spacing is much larger than particle size. Then the
agglomerates are elongated and tree-like even at the beginning of the aggregation process. This is particularly
true for small agglomerates as confirmed by the small value of the coordination index and the larger eccentricity
index. These indices give a more complete description of the agglomeration process than the fractal exponent
and its prefactor \cite{Hein} alone. In fact, these indices provide information about the local connectivity
and mass distribution inside the agglomerate. Their behavior in terms of agglomerate size is opposite, the
average coordination (eccentricity) index increases (decreases) with agglomerate size so that the geometric
mean of both indices is roughly constant with agglomerate size.\\

The main achievements of our work can be recapitulated as follows:
\begin{itemize}
\item The fractal exponent is not a fixed value determined by the kind of aggregation process
(DLA or DLCA) that has taken place. Instead the fractal exponent is closely related to the density of primary particles that will agglomerate. We base this assertion on Monte Carlo simulation results carried out in real physical space.
\item The aggregates are characterized by the fractal exponent and by two other geometric parameters, the coordination and eccentricity indices. The behaviors of these indices reinforce the conclusion that aggregates become self-similar for large times if their size is sufficient. The geometric mean of the coordination and eccentricity indices is almost the same for different primary densities (and therefore for different fractal exponents) which suggest that these indices are related once self-similarity has set in.
\end{itemize}

%\textbf{Experiments are lacking for the moment and has to constitute the natural continuation of the present work. / D-R2}.
Although our simulations refer to particle agglomeration during combustion, the simulation algorithm is applicable to many other agglomeration processes. In particular, we may also generalise the algorithm to include thermophoretic forces over the agglomerates during agglomeration. We are currently working in this direction.\\

\begin{acknowledgments}
We thank Manuel Arias Zugasti from UNED for fruitful discussions and useful suggestions. This work has been
supported by the Spanish Ministerio de Econom\'\i a y Competitividad grant FIS2011-28838-C02-01 and by the
Autonomous Region of Madrid grant P2009/ENE-1597 (HYSYCOMB).
\end{acknowledgments}

\section*{Appendix: Physical time equivalence}
To establish the simulation time step according to (\ref{time_step}), we need the time-dependent mean squared
displacement of a particle, $\langle\Delta x^2\rangle$. For the Ornstein-Uhlenbeck velocity autocorrelation,
(\ref{time_step}) is
\begin{eqnarray}
\langle\Delta x^2\rangle&=&2D_{p}\displaystyle{\left\{t+\varTheta\left[\exp\left(-\frac{t}{\varTheta}\right)-1\right]\right\}}\nonumber\\[1ex]
\varTheta&=&\displaystyle{\frac{m}{\zeta}}\,,\quad\quad\quad\text{particle relaxation time}\nonumber\\[1 ex]D_{p}&=&\displaystyle{\frac{k_{B}T}
{\zeta}}\,,\quad\quad\text{particle diffusion coefficient}\nonumber\\[1ex]
\zeta&=&\displaystyle{\frac{3\,\pi\,\mu_{a}\,d_{p}}{C}}\,,\quad\quad\text{particle friction coefficient}\nonumber\\[1ex]
C&=&1+\displaystyle{\frac{2\,l_{a}}{d_{p}}\left[1.257+0.4\exp\left(-\frac{0.55\,d_{p}}{l_{a}}\right)\right]}\,,\nonumber\\[2ex]
l_{a}&=&\displaystyle{\frac{1}{\sqrt{2}\,\pi\,d_{a}^2\,n_{a}}}\,,\quad\text{mean free path of air molecules,}
\nonumber
\end{eqnarray}
where $C$ is the slip correction factor, $m$ is the particle mass, $T$ is the fluid (air) temperature, $d_{a}$ is the fluid (air) mean molecular
diameter, $n_{a}$ is the fluid (air) number density and $\mu_{a}$ is the fluid (air) viscosity.

\begin{figure}[b]
\includegraphics[width=10cm,height=5cm]{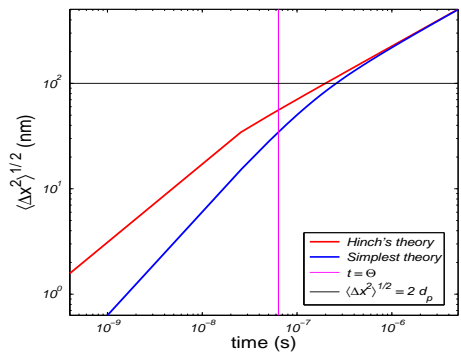}
\caption{\label{msd_t} \textit{Mean-square displacement as a function of time according to Hinch's and the simplest theory of Brownian motion. Time step
corresponding to the adopted jump length of $2\,d_{p}$ ($10^2$) is greater than the particle relaxation time ($\Theta$), as indicated.}}
\end{figure}

Hinch's theory of Brownian motion takes the hydrodynamic interactions between particles and fluid into account,
and it produces the following mean squared displacement \cite{wei89}:
\begin{widetext}
\begin{eqnarray}
\langle\Delta x^2\rangle&=&2D_{p}\displaystyle{\left\{t-2\sqrt{\frac{\tau\,t}{\pi}}+\frac{2\tau}{9}\left(1-\frac{\rho_{p}}{\rho_{a}}\right)
+\frac{3}{\sqrt{\tau(5-8\rho_{p}/\rho_{a})}}\left[\frac{1}{a_{+}^3}\mathrm{e}^{a_{+}^2t}\mathrm{erfc}(a_{+}\sqrt{t})-\frac{1}{a_{-}^3}
\mathrm{e}^{a_{-}^2t}\mathrm{erfc}(a_{-}\sqrt{t})\right]\right\}}\nonumber\\[1ex]
a_{\pm}&=&\displaystyle{\frac{3}{2}\left[\frac{3\pm\sqrt{5-8\rho_{p}/\rho_{a}}}{\sqrt{\tau}(1+2\rho_{p}/\rho_{a})}\right]}\nonumber\\[1ex]
\tau&=&\displaystyle{\frac{d_{p}^2\,\rho_{a}}{4\,\mu_{a}}},\qquad\text{time for diffusion of vorticity across a particle radius}\nonumber
\end{eqnarray}
\end{widetext}
where $\rho_{p}$ is the particle mass density, and $\rho_{a}$ is the fluid (air) mass density. The slip correction
factor makes these expressions to be valid for both the continuum and the free molecular regimes. The time
corresponding to a jump length of $2\,d_{p}$, according with the Hinch's theory is $2\times10^{-7}$ sec.
approximately, which is larger than the particle relaxation time as it can be seen in Figure \ref{msd_t}.

\bibliography{paper}

\end{document}